\documentclass[times]{GonnellaLab-StyleArXiv}

\usepackage{blindtext}
\usepackage[overload]{textcase}
\usepackage{amsmath,amssymb}
\usepackage{lmodern}
\usepackage{iftex}
\usepackage{enumitem}
\usepackage{pifont}
\usepackage{booktabs}
\usepackage{multicol}
\usepackage[T1]{fontenc}
\usepackage[utf8]{inputenc}
\usepackage{textcomp} 
\usepackage{color}
\usepackage{fancyvrb}
\usepackage{xcolor}
\usepackage{natbib}

\IfFileExists{upquote.sty}{\usepackage{upquote}}{}
\IfFileExists{microtype.sty}{
  \usepackage[]{microtype}
  \UseMicrotypeSet[protrusion]{basicmath} 
}{}
\makeatletter
\@ifundefined{KOMAClassName}{
  \IfFileExists{parskip.sty}{%
    \usepackage{parskip}
  }{
    \setlength{\parindent}{0pt}
    \setlength{\parskip}{6pt plus 2pt minus 1pt}}
}{
  \KOMAoptions{parskip=half}}
\makeatother
\IfFileExists{xurl.sty}{\usepackage{xurl}}{} 
\IfFileExists{bookmark.sty}{\usepackage{bookmark}}{}
\hypersetup{
  hidelinks,
  pdfcreator={LaTeX via pandoc}}
\DefineVerbatimEnvironment{Highlighting}{Verbatim}{commandchars=\\\{\}}
\newenvironment{Shaded}{}{}
\newcommand{\SectionTok}[1]{\textcolor[rgb]{0.02,0.04,0.15}{#1}}
\newcommand{\DtnameTok}[1]{\textcolor[rgb]{0.65,0.37,0.50}{#1}}
\newcommand{\DefkeyTok}[1]{\textcolor[rgb]{0.49,0.56,0.16}{#1}}
\newcommand{\OptionTok}[1]{\textcolor[rgb]{0.30,0.56,0.16}{\textit{#1}}}

\newcommand{\AttributeTok}[1]{\textcolor[rgb]{0.49,0.56,0.16}{#1}}
\newcommand{\CharTok}[1]{\textcolor[rgb]{0.25,0.44,0.63}{#1}}
\newcommand{\CommentTok}[1]{\textcolor[rgb]{0.38,0.63,0.69}{\textit{#1}}}
\newcommand{\DecValTok}[1]{\textcolor[rgb]{0.25,0.63,0.44}{#1}}
\newcommand{\FunctionTok}[1]{\textcolor[rgb]{0.02,0.16,0.49}{#1}}
\newcommand{\KeywordTok}[1]{\textcolor[rgb]{0.00,0.44,0.13}{\textbf{#1}}}

\newcommand{\SpecialCharTok}[1]{\textcolor[rgb]{0.25,0.44,0.63}{#1}}
\newcommand{\StringTok}[1]{\textcolor[rgb]{0.25,0.44,0.63}{#1}}
\newcommand{\VariableTok}[1]{\textcolor[rgb]{0.10,0.09,0.49}{#1}}
\ifLuaTeX
  \usepackage{selnolig}  
\fi

\begin{document}
    
\leadauthor{Gonnella}


\title{Designing text representations for existing
data using the TextFormats Specification Language}
\shorttitle{TFSL definitions by value type}

\author[1,2\space \Letter]{Giorgio Gonnella}

\affil[1]{Center for Bioinformatics (ZBH), Universität Hamburg, Bundesstrasse 43, 20146 Hamburg}
\affil[2]{Institute for Microbiology and Genetics, Georg-August-Universität Göttingen, Goldschmidtstr. 1, 37077 Göttingen}

\maketitle

\begin{abstract}
TextFormats is a software system for efficient and user-friendly creation of text format specifications, accessible from multiple programming languages (C/C++, Python, Nim) and the Unix command line. To work with a format, a specification written in the TextFormats Specification Language (TFSL) must be created. The specification defines datatypes for each part of the format.

The syntax for datatype definitions in TextFormats specifications is based on the text representation. Thus this system is well suited for the description of existing formats.
However, when creating a new text format
for representing existing data, the user may use different possible definitions,
based on the type of value and the representation choices.

This study explores the
possible definition syntax in the TextFormats Specification Language to be used for
creating text representations of scalar values (e.g.\ string, numeric value, boolean) and compound data structures (e.g.\ array, mapping).
The results of the analysis are presented systematically, together with examples for each 
each type of different values that can be represented, and usage advices.

\end {abstract}


\begin{corrauthor}
    giorgio.gonnella\at uni-goettingen.de
\end{corrauthor}

\begin{multicols}{2}


TextFormats \citep{textformats} is a software library and a set of tools for defining text formats. Although it was initially developed for the representation of bioinformatics formats, it is a generic software which can be applied to a variety of fields. Once a format has been defined using the
domain-specific language TFSL (TextFormats Specification Language), the
TextFormats specification can be used for parsing the format, as well as writing data in the format, using the APIs for several programming languages
(Nim, Python, C/C++) or in scripts, using the provided Unix command-line tools.

In TFSL, various kinds of definitions are used to describe different types of data textual representations, as shown in Table \ref{tab:DefKeys}. These definitions are dependent not only on the type of value but also on other characteristics of the data and its representation. For example they depend on the set of possible values: e.g. for a string, one uses \texttt{constant} for a single value, \texttt{values} for a small set of values and \texttt{regex} for all strings matching a regular expression. Also different kind of definitions can sometimes be used for the same set of strings (e.g. \texttt{values}, \texttt{regex} and \texttt{regexes}). For some compound data values, the kind of definition to use depends on if and how semantic and datatype of the composing elements is coded in the text representation (e.g. \texttt{composed\_of} vs.\ \texttt{labeled\_list} or \texttt{tagged\_list}). Thus it is interesting to investigate the use cases of these different definition kinds, depending on the type of represented value.

It is important to note that there is not always a one-to-one correspondence between the type of represented value and its text representation. For instance, the text representation "I" could represent the string "I" (e.g. as the first person singular pronoun in English) or represent the integer value expressed as a Roman numeral, which is more commonly represented using Arabic numerals (``1'').

This paper presents a systematic exploration of the different types of values that can be represented in portions of a text format and reports the corresponding TextFormats definitions required to accurately describe the data in a specification. For each type of value it includes practical examples of TFSL definitions to illustrate the concepts presented.


\begin{table*}[!t]
\centering
\begin{tabular}{ll}
\toprule
\textbf{Definition key} & \textbf{Description}\\
\midrule
& \\[-3mm]
\textit{for scalar data} & \\[1mm]
\ \ \ \ \texttt{constant} & Single, invariant value \\
\ \ \ \ \texttt{values}   & One of a set of values \\
\ \ \ \ \texttt{regex}   & Matches provided regular expression \\
\ \ \ \ \texttt{regexes} & Matches one of the provided regular expressions \\
\ \ \ \ \texttt{integer}  & Signed integer, optionally with range validation \\
\ \ \ \ \texttt{unsigned\_integer} & Unsigned integer, optionally with range validation \\
\ \ \ \ \texttt{float} & Floating point value, optionally with range validation \\
& \\[-3mm]
\textit{for compound data} & \\[1mm]
\ \ \ \ \texttt{list\_of} &  Ordered set; element datatype and semantic independent of position \\
\ \ \ \ \texttt{composed\_of} & Ordered set; element datatype and semantic dependent of position \\
\ \ \ \ \texttt{labeled\_list} & Key/value pairs set; element datatype and semantic dependent on the key \\
\ \ \ \ \texttt{tagged\_list} & Tagname/typecode/value triples; semantic dependent on tagname, datatype on typecode \\
& \\[-1.5mm]
\textit{multiple choices}\\[1mm]
\ \ \ \ \texttt{one\_of} & One of different formats, described by separate definitions \\
& \\[-3mm]
\bottomrule
\end{tabular}
\caption{Datatype definition keys available in the TextFormats Specification Language, according to the class of data type to be defined (scalar or compound), as well as for the definition of multiple choices for a single element (\texttt{one\_of}).}
\label{tab:DefKeys}
\vspace{5mm}
\end{table*}


\section*{Types of Data Values}

In the context of text formats, various type of values can be represented. One important distinction to consider is whether a data point logically represents an indivisible, atomic unit or can be broken down into multiple components. If it is the former, it is considered a \textit{scalar} value, and if it is the latter, it is considered a \textit{compound} value.

Scalar values are single, atomic units of data. This category encompasses several subtypes, including numerical data, categorical data, boolean values, and  strings. Numerical data can include integer and floating-point values, and is used to represent mathematical values. Categorical data, on the other hand, assigns each data point to a predefined, finite set of possible values. Boolean values are an example of categorical data, for representing a binary condition, and can be either true or false. Finally, strings are sequences of characters, and can be used to represent a variety of values.

Compound values consist of multiple components, each of which can be a scalar or itself also a compound value. Compound data types provide a way to organize data into a complex structure, allowing for the representation of structured, multi-layered information. For instance, a compound data type could consist of several numerical values, each representing a different data point, or it could consist of multiple components, each representing a different aspect of the data. The important feature of compound data types is that they allow for the representation of multi-faceted information in a way that can be parsed and analyzed in a meaningful manner. Thus the semantics and data type of each component must be known to the parser, either as an external convention, or through metadata included in the text representation itself (e.g. tags).

\section*{Definitions for Scalar Values}
\subsection*{Numeric values}

Several kind of datatype definitions are dedicated to numerical variables.
In order to define numerical values, the following criteria are to be considered.
First, if the value is an integer or not. Second, what is the range of the
possible values. Third, because they are internally represented differently in many contexts (e.g. C), if a value is an integer, it shall be considered if it is
signed or not.

\subsubsection*{Any value}
If every valid value is acceptable, as long as it fits in the range of the data type in the programming language or environment in which the data is used, then
the predefined \texttt{integer}, \texttt{unsigned\_integer} and
\texttt{float} are used. Since they are predefined, they do not consist in
a definition, but the key (e.g. \texttt{integer}) is simply inserted in
the specification in the position where the definition would be required.
This is done usually, in a compound datatype or an alias, as in the following examples:

\begin{Shaded}
\begin{Highlighting}[]
\SectionTok{datatypes}\KeywordTok{:}
\AttributeTok{  }\DtnameTok{num1}\KeywordTok{:}\AttributeTok{ }\DefkeyTok{unsigned\_integer}
\AttributeTok{  }\DtnameTok{list1}\KeywordTok{:}\AttributeTok{ }\KeywordTok{\{}\DefkeyTok{list\_of}\KeywordTok{:}\DefkeyTok{ integer}\KeywordTok{\}}
\end{Highlighting}
\end{Shaded}

\subsubsection*{Values in a range}

If only values in a given range are accepted, the options \texttt{min} and/or
\texttt{max} can be used:

\begin{Shaded}
\begin{Highlighting}[]
\SectionTok{datatypes}\KeywordTok{:}
\AttributeTok{  }\DtnameTok{num2}\KeywordTok{:}\DefkeyTok{ integer: }\VariableTok{\{}\OptionTok{min}\VariableTok{: }\DecValTok{{-}1}\VariableTok{, \OptionTok{max}: }\DecValTok{1}\VariableTok{\}}
\AttributeTok{  }\DtnameTok{num3}\KeywordTok{:}\DefkeyTok{ unsigned\_integer: }
\VariableTok{                  \{\OptionTok{min}: }\DecValTok{0}\VariableTok{, \OptionTok{max}: }\DecValTok{100}\VariableTok{\}}
\AttributeTok{  }\DtnameTok{num4}\KeywordTok{:}\DefkeyTok{ float: }\VariableTok{\{\OptionTok{min}: \DecValTok{0}, \OptionTok{max}: \DecValTok{1}\}}
\end{Highlighting}
\end{Shaded}

The default is to include the limits in the
accepted values, but these can be excluded using
\texttt{(min\textbar{}max)\_excluded}. This is mostly useful only for floating
point values, since for integers it suffices to increase or decrease the limit
by 1.

\begin{Shaded}
\begin{Highlighting}[]
\SectionTok{datatypes}\KeywordTok{:}
\AttributeTok{  }\DtnameTok{num5}\KeywordTok{:}\DefkeyTok{ float: }\VariableTok{\{\OptionTok{min}: \DecValTok{0}, \OptionTok{max}: \DecValTok{100},}
\VariableTok{                \OptionTok{min\_excluded}: \CharTok{true},}
\VariableTok{                \OptionTok{max\_excluded}: \CharTok{true}\}}
\end{Highlighting}
\end{Shaded}

\subsubsection*{Element of a set of values}
If only elements of a given set of values shall be representable, the \texttt{values}
definition key is used, as in the following example:

\begin{Shaded}
\begin{Highlighting}[]
\SectionTok{datatypes}\KeywordTok{:}
\AttributeTok{  }\DtnameTok{num6}\KeywordTok{:}\AttributeTok{ }\KeywordTok{\{}\DefkeyTok{values}\KeywordTok{:}\AttributeTok{ }\KeywordTok{[}\DecValTok{1}\KeywordTok{,}\DecValTok{2}\KeywordTok{,}\DecValTok{3}\KeywordTok{]\}}
\end{Highlighting}
\end{Shaded}

\vfill\null\pagebreak

A limitation is that since all possible values must be enumerated,  the performance
of this definition is very bad, if the number of elements of the set is
too high.

\subsubsection*{Single possible value}
If only a single value shall be
accepted a \texttt{constant}
definition is used:

\begin{Shaded}
\begin{Highlighting}[]
\SectionTok{datatypes}\KeywordTok{:}
\AttributeTok{  }\DtnameTok{num7}\KeywordTok{:}\AttributeTok{ }\KeywordTok{\{}\DefkeyTok{constant}\KeywordTok{:}\AttributeTok{ }\DecValTok{3}\KeywordTok{\}}
\end{Highlighting}
\end{Shaded}

\subsubsection*{Roman numerals}

The support in TFSL of numerical values in non-conventional representations is
very limited. The only case which can be represented with ease is the case in
which all possible values can be enumerated - and thus their number is limited,
otherwise the performance would be negatively affected. In this case
a \texttt{values} definition can be used, by specifying conversions for each
element to the corresponding numerical value.
An example is given here, for representing the roman numbers from 1 to 3:

\begin{Shaded}
\begin{Highlighting}[]
\SectionTok{datatypes}\KeywordTok{:}
\AttributeTok{  }\DtnameTok{num8}\KeywordTok{:}
\AttributeTok{    }\DefkeyTok{values}\KeywordTok{:}
\AttributeTok{      }\KeywordTok{{-}}\AttributeTok{ }\StringTok{"I"}\KeywordTok{:}\AttributeTok{ }\DecValTok{1}
\AttributeTok{      }\KeywordTok{{-}}\AttributeTok{ }\StringTok{"II"}\KeywordTok{:}\AttributeTok{ }\DecValTok{2}
\AttributeTok{      }\KeywordTok{{-}}\AttributeTok{ }\StringTok{"III"}\KeywordTok{:}\AttributeTok{ }\DecValTok{3}
\end{Highlighting}
\end{Shaded}

In other cases (such as if any value in a range or all numeric value must be representable), the parsing or formatting of the values cannot be directly
defined in TextFormats. Thus the data type must be defined as a string (e.g. by a regex) and handled by the calling code, as in the following example:

\begin{Shaded}
\begin{Highlighting}[]
\SectionTok{datatypes}\KeywordTok{:}
\AttributeTok{  }\DtnameTok{num9}\KeywordTok{:}\AttributeTok{ }\KeywordTok{\{}\DefkeyTok{regex}\KeywordTok{:}\AttributeTok{ }\StringTok{"^[MDCLXVI]+\$"}\KeywordTok{\}}
\end{Highlighting}
\end{Shaded}


\hypertarget{boolean-and-undefined-values}{%
\subsection{Boolean variables}\label{boolean-and-undefined-values}}
Booleans are variables that can only contain one of two values: true or false.
Their representation is usually a pair of strings such as "T" and "F",
or "0" and "1".

\subsubsection*{Single representations}
The following definition shows how to define a type for a boolean variable,
with a string value for each of the two decoded values (true or false).
These representations can be provided using an
\texttt{values} definition and a mapping:

\begin{Shaded}
\begin{Highlighting}[]
\SectionTok{datatypes}\KeywordTok{:}
\AttributeTok{  }\DtnameTok{boolean1}\KeywordTok{:}
\AttributeTok{    }\DefkeyTok{values}\KeywordTok{:}\AttributeTok{ }\KeywordTok{[}\FunctionTok{"T"}\KeywordTok{:}\AttributeTok{ }\CharTok{true}\KeywordTok{,}\AttributeTok{ }\FunctionTok{"F"}\KeywordTok{:}\AttributeTok{ }\CharTok{false}\KeywordTok{\}}\AttributeTok{ }
\end{Highlighting}
\end{Shaded}

\subsubsection*{Multiple representations}

Sometimes multiple string representations are accepted in a format for
each of the two values of a boolean. In this case a \texttt{values}
or \texttt{regex} definition is used and canonical representations must be specified: 

\begin{Shaded}
\begin{Highlighting}[]
\SectionTok{datatypes}\KeywordTok{:}
\AttributeTok{  }\DtnameTok{boolean2}\KeywordTok{:}
\AttributeTok{    }\DefkeyTok{regexes}\KeywordTok{:}
\AttributeTok{      }\KeywordTok{{-}}\AttributeTok{ }\StringTok{"[Tt](rue)?"}\KeywordTok{:}\AttributeTok{ }\CharTok{true}
\AttributeTok{      }\KeywordTok{{-}}\AttributeTok{ }\StringTok{"[Ff](alse)?"}\KeywordTok{:}\AttributeTok{ }\CharTok{false}
\AttributeTok{    }\OptionTok{canonical}\KeywordTok{:}\AttributeTok{ }\KeywordTok{\{}\FunctionTok{"T"}\KeywordTok{:}\AttributeTok{ }\CharTok{true}\KeywordTok{,}\AttributeTok{ }\FunctionTok{"F"}\KeywordTok{:}\AttributeTok{ }\CharTok{false}\KeywordTok{\}}
\end{Highlighting}
\end{Shaded}

\subsubsection*{Presence or absence}
In other cases, the value of a boolean variable is encoded as the presence or absence of
a given element in the text representation. In this case a \texttt{constant} definition with a
default value (\texttt{empty} key) can be used:

\begin{Shaded}
\begin{Highlighting}[]
\SectionTok{datatypes}\KeywordTok{:}
\AttributeTok{  }\DtnameTok{boolean3}\KeywordTok{:}
\AttributeTok{    }\DefkeyTok{constant}\KeywordTok{:}\AttributeTok{ }\KeywordTok{\{}\FunctionTok{"\$"}\KeywordTok{:}\AttributeTok{ }\CharTok{true}\KeywordTok{\}}
\AttributeTok{    }\OptionTok{empty}\KeywordTok{:}\AttributeTok{ }\CharTok{false}
\end{Highlighting}
\end{Shaded}

\subsubsection*{Handling the undefined state}

In some cases, a three-way variable state is possible.
For example, a variable could take
a boolean value or a special \textit{undefined} 
value. In such cases, further options are simply
added to the \texttt{values} definition mapping:

\begin{Shaded}
\begin{Highlighting}[]
\SectionTok{datatypes}\KeywordTok{:}
\AttributeTok{  }\DtnameTok{boolean4}\KeywordTok{:}
\AttributeTok{    }\DefkeyTok{values}\KeywordTok{:}\AttributeTok{ }\KeywordTok{[}\FunctionTok{"T"}\KeywordTok{:}\AttributeTok{ }\CharTok{true}\KeywordTok{,}\AttributeTok{ }\FunctionTok{"F"}\KeywordTok{:}\AttributeTok{ }\CharTok{false}\KeywordTok{,}
\AttributeTok{             }\FunctionTok{"NA"}\KeywordTok{:}\AttributeTok{ }\CharTok{null}\KeywordTok{]}
\end{Highlighting}
\end{Shaded}

\hypertarget{strings}{%
\subsection{String values}\label{strings}}

The present section describe how to define the data types for values,
which are stored in string variables.

\subsubsection*{Any string}

If a component of a format can contain any string, the predefined
\texttt{string} datatype can be used for it. This is used in aliases
or compound definitions:

\begin{Shaded}
\begin{Highlighting}[]
\SectionTok{datatypes}\KeywordTok{:}
\AttributeTok{  }\DtnameTok{str1}\KeywordTok{:}\AttributeTok{ }\DefkeyTok{string}
\AttributeTok{  }\DtnameTok{list2}\KeywordTok{:}\AttributeTok{ }\KeywordTok{\{}\DefkeyTok{list\_of}\KeywordTok{:}\DefkeyTok{ string}\KeywordTok{\}}
\end{Highlighting}
\end{Shaded}

\subsubsection*{Categorical labels}

Categorical variables have values, which can only take a limited number
of different values, which can be enumerated. Some cases of categorical
values have already
been handled above, i.e. boolean values and numerical variables,
with a fixed number of possible values. These were defined using
a \texttt{values} definition.
Also if the categories are stored in strings, the same kind of
definition is used:

\begin{Shaded}
\begin{Highlighting}[]
\SectionTok{datatypes}\KeywordTok{:}
\AttributeTok{  }\DtnameTok{str2}\KeywordTok{:}\AttributeTok{ }\KeywordTok{\{}\DefkeyTok{values}\KeywordTok{:}\AttributeTok{ }\KeywordTok{[}\StringTok{"ABC"}\KeywordTok{,}\AttributeTok{ }\StringTok{"DEF"}\KeywordTok{]\}}
\end{Highlighting}
\end{Shaded}

\subsubsection*{Single possible value}
If a string variable may only contain a single value, then
a \texttt{constant} definition is used:

\begin{Shaded}
\begin{Highlighting}[]
\SectionTok{datatypes}\KeywordTok{:}
\AttributeTok{  }\DtnameTok{str3}\KeywordTok{:}\AttributeTok{ }\KeywordTok{\{}\DefkeyTok{constant}\KeywordTok{:}\AttributeTok{ }\StringTok{"XYZ"}\KeywordTok{\}}
\end{Highlighting}
\end{Shaded}

\subsubsection*{Regular expressions}

If a string must match a given regular expression, a definition of kind
\texttt{regex} is used:

\begin{Shaded}
\begin{Highlighting}[]
\SectionTok{datatypes}\KeywordTok{:}
\AttributeTok{  }\DtnameTok{str4}\KeywordTok{:}\AttributeTok{ }\KeywordTok{\{}\DefkeyTok{regex}\KeywordTok{:}\AttributeTok{ }\StringTok{"[ABC]\{1,3\}"}\KeywordTok{\}}
\end{Highlighting}
\end{Shaded}

In some cases, it is easier to split a regex into multiple pieces,
which are defined separately. While the universe of matching strings
remains the same, if a single or multiple regexes are used, the
use of multiple regexes has the advantage that the single regexes could
be more readable, but also that each piece can be
treated differently, when the strings are mapped to given values:

\begin{Shaded}
\begin{Highlighting}[]
\SectionTok{datatypes}\KeywordTok{:}
\AttributeTok{  }\DtnameTok{str5}\KeywordTok{:}
\AttributeTok{    }\DefkeyTok{regexes}\KeywordTok{:}
\AttributeTok{    }\KeywordTok{{-}}\AttributeTok{ }\StringTok{"[ABC]\{1,3\}"}
\AttributeTok{    }\KeywordTok{{-}}\AttributeTok{ }\StringTok{"[DEF]\{5,7\}"}
\end{Highlighting}
\end{Shaded}

\subsubsection*{Empty strings}

If the empty string shall also be handled, its value can be provided using
\texttt{empty} (this option is available for any kind of definition).
The option has the highest
priority, thus also if e.g.~a regular expression matching also an empty
string is provided, the empty string case is handled as defined in the \texttt{empty} option. E.g.\ in the following case, the empty string results in a decoded value 0:

\begin{Shaded}
\begin{Highlighting}[]
\SectionTok{datatypes}\KeywordTok{:}
\AttributeTok{  }\DtnameTok{str6}\KeywordTok{:} \KeywordTok{\{}\DefkeyTok{regex}\KeywordTok{:}\AttributeTok{ }\StringTok{"\textbackslash{}d*"}\AttributeTok{, }\OptionTok{empty}\KeywordTok{:}\AttributeTok{ }\StringTok{"0"}\KeywordTok{\}}
\end{Highlighting}
\end{Shaded}

\hypertarget{validation-of-a-string-by-datatype-definition}{%
\subsubsection{Structured strings}\label{validation-of-a-string-by-datatype-definition}}

In some cases, although a value shall be decoded as string, and not
further parsed into smaller elements, it has an internal structure.
In this case it can be useful to create a definition (e.g.~for a
compound datatype, as explained below) and then let TextFormats know
that the definition shall only be used for validation, but not for
parsing, using the \texttt{as\_string:\ true} option.

For example the following definition of a string (containing unsigned
integers and `.' separating them) uses a relatively complex regular
expression:

\begin{Shaded}
\begin{Highlighting}[]
\SectionTok{datatypes}\KeywordTok{:}
\AttributeTok{  }\DtnameTok{str7}\KeywordTok{:}
\AttributeTok{    }\DefkeyTok{regex}\KeywordTok{:}\AttributeTok{ }\StringTok{"(0|[1{-}9][0{-}9]*)}
\StringTok{            (\textbackslash{}.(0|[1{-}9][0{-}9]*))*"}
\end{Highlighting}
\end{Shaded}

The following definition is equivalent, but more readable:

\begin{Shaded}
\begin{Highlighting}[]
\SectionTok{datatypes}\KeywordTok{:}
\AttributeTok{  }\DtnameTok{str8}\KeywordTok{:}
\AttributeTok{    }\DefkeyTok{list\_of}\KeywordTok{:}\AttributeTok{ }\KeywordTok{\{}\DefkeyTok{regex}\KeywordTok{:}\AttributeTok{ unsigned\_integer}\KeywordTok{\}}
\AttributeTok{    }\OptionTok{splitted\_by}\KeywordTok{:}\AttributeTok{ }\StringTok{"."}
\AttributeTok{    }\OptionTok{min\_length}\KeywordTok{:}\AttributeTok{ }\DecValTok{1}
\AttributeTok{    }\OptionTok{as\_string}\KeywordTok{:}\AttributeTok{ }\CharTok{true}
\end{Highlighting}
\end{Shaded}

\subsubsection*{Acronyms}

By default, string are just encoded as the string itself, i.e.~parsing
involves validation, but no modification of the value itself. In some
cases a different string should be present in the string representation
compared to the decoded value: e.g.~one could want to expand an acronym.
For this a decoded mapping is used:

\begin{Shaded}
\begin{Highlighting}[]
\SectionTok{datatypes}\KeywordTok{:}
\AttributeTok{  }\DtnameTok{str9}\KeywordTok{:}
\AttributeTok{    }\DefkeyTok{values}\KeywordTok{:}
\AttributeTok{      }\KeywordTok{{-}}\AttributeTok{ }\StringTok{"USA"}\KeywordTok{:}\AttributeTok{ }\StringTok{"United States"}
\AttributeTok{      }\KeywordTok{{-}}\AttributeTok{ }\StringTok{"UK"}\KeywordTok{:}\AttributeTok{ }\StringTok{"United Kingdom"}
\end{Highlighting}
\end{Shaded}

\vfill\null\columnbreak

If multiple encoded values are decoded to the same decoded value (always
for regular expressions), then canonical encoded forms must be
specified, so that the encoder knows which one shall be used. E.g.:
\begin{Shaded}
\begin{Highlighting}[]
\SectionTok{datatypes}\KeywordTok{:}
\AttributeTok{  }\DtnameTok{str10}\KeywordTok{:}
\AttributeTok{    }\DefkeyTok{regexes}\KeywordTok{:}
\AttributeTok{      }\KeywordTok{{-}}\AttributeTok{ }\StringTok{"U(SA?|sa)"}\KeywordTok{:}\AttributeTok{ }\StringTok{"United States"}
\AttributeTok{      }\KeywordTok{{-}}\AttributeTok{ }\StringTok{"U[Kk]"}\KeywordTok{:}\AttributeTok{ }\StringTok{"United Kingdom"}
\AttributeTok{    }\OptionTok{canonical}\KeywordTok{:}
\AttributeTok{      }\KeywordTok{{-}}\AttributeTok{ }\StringTok{"USA"}\KeywordTok{:}\AttributeTok{ }\StringTok{"United States"}
\AttributeTok{      }\KeywordTok{{-}}\AttributeTok{ }\StringTok{"UK"}\KeywordTok{:}\AttributeTok{ }\StringTok{"United Kingdom"}
\end{Highlighting}
\end{Shaded}

\section*{Definitions for Compound Values}

There are several different types of compound data.
Hereby we distinguish 3 cases, which are handled separately:
\begin{enumerate}
\item ordered lists of elements, which are semantically equivalent
\item dictionaries or mappings, i.e.\ open associative data structures in which different elements may have distinct semantics
\item objects or structures, i.e.\ which generally contain different, semantically distinct attributes
\end{enumerate}

It is worth noticing that the names of the data types for these kinds of compound values depend on the context, such as programming language, and on the underlying data structure, i.e.\ how the data is stored in memory.

\subsubsection*{Compound Data Compatibility in TextFormats}

Compound values can include other compound values as components. This hierarchical structure can be represented using a tree, with the depth potentially being indefinite and marked in the text representation using indentation or nested pairs of parentheses. However, for a format to be compatible with the current implementation of TextFormats, it must represent a regular language (with the possible exception of parts of the format handled by an external library, i.e.\ currently JSON). As a result, the definition tree in compound data types must be known at definition time and circular definitions are not allowed in TFSL.

\subsection*{Lists, Arrays, Sequences, Sets}

We consider here compound data values, where the elements are ordered (the order may, but must not, be meaningful), but they are all considered semantically equivalent and the element data type and set of representable values is not dependent on their position in the list.

\vfill\null\pagebreak

If the order matters, the compound values are often stored in an array or list data structure (e.g. a linked list) depending
on the underlying data structure and the context (e.g. programming language), and are thus called \textit{lists} (e.g. Python), \textit{arrays} (e.g. C, Python) or \textit{sequences} (e.g. Nim, YAML). Values of this kind are described in TextFormats using \texttt{list\_of} datatype definitions. 

The definition of compound values where the semantics of the elements does not differ among the elements is done in TextFormats using the \texttt{list\_of} key, as in the following example:

\begin{Shaded}
\begin{Highlighting}[]
\SectionTok{datatypes}\KeywordTok{:}
\AttributeTok{  }\DtnameTok{list3}\KeywordTok{:}
\AttributeTok{    }\DefkeyTok{list\_of}\KeywordTok{:}\AttributeTok{ }\KeywordTok{\{}\DefkeyTok{regex}\KeywordTok{:}\AttributeTok{ }\KeywordTok{[}\AttributeTok{A{-}Z}\KeywordTok{]\}}
\end{Highlighting}
\end{Shaded}

\subsubsection*{Element separators}

Often the parsing of the single elements is made possible by a separator string between the
elements, which does not occur in the elements itself. This is specified
using the option \texttt{splitted\_by}:

\begin{Shaded}
\begin{Highlighting}[]
\SectionTok{datatypes}\KeywordTok{:}
\AttributeTok{  }\DtnameTok{list4}\KeywordTok{:}
\AttributeTok{    }\DefkeyTok{list\_of}\KeywordTok{:}\AttributeTok{ unsigned\_integer}
\AttributeTok{    }\OptionTok{splitted\_by}\KeywordTok{:}\AttributeTok{ }\StringTok{","}
\end{Highlighting}
\end{Shaded}

\subsubsection*{Separator escaping}

In some cases, however, the separating string can also be present in the
elements itself, e.g.~by escaping it. In this case the
\texttt{separator} option is used in the definition,
and a e.g.~a regular expression is used for defining the elements:

\begin{Shaded}
\begin{Highlighting}[]
\SectionTok{datatypes}\KeywordTok{:}
\AttributeTok{  }\DtnameTok{list5}\KeywordTok{:}
\AttributeTok{    }\DefkeyTok{list\_of}\KeywordTok{:}
\AttributeTok{      }\DefkeyTok{regex}\KeywordTok{:}\AttributeTok{ }\StringTok{"(}\SpecialCharTok{\textbackslash{}\textbackslash{}}\StringTok{\textbackslash{}:|[A{-}Za{-}z0{-}9 \_])*"}
\CommentTok{        \# allows : escaped by \textbackslash{}}
\AttributeTok{    }\OptionTok{separator}\KeywordTok{:}\AttributeTok{ }\StringTok{":"}
\end{Highlighting}
\end{Shaded}

Given the definition above, for example, the string\\
\texttt{elem 1:elem2:elem\_3:elem\textbackslash{}:\textbackslash{}:4}\\
would be parsed into the four elements
\texttt{elem1}, \texttt{elem2}, \texttt{elem\_3} and
\texttt{elem\textbackslash{}:\textbackslash{}:4}.

\subsubsection*{Fixed length elements}

In case the elements of a list have all the same length, a separator is generally not necessary. However, if one is present, it may be also present in the element
text itself, since there is no risk of confusion. In this case the 
\texttt{separator} option is used, e.g.:

\begin{Shaded}
\begin{Highlighting}[]
\SectionTok{datatypes}\KeywordTok{:}
\AttributeTok{  }\DtnameTok{list6}\KeywordTok{:}
\AttributeTok{    }\DefkeyTok{list\_of}\KeywordTok{:}
\AttributeTok{      }\DefkeyTok{regex}\KeywordTok{:}\AttributeTok{ }\StringTok{"[:0{-}9]\{3\}"}
\AttributeTok{      \OptionTok{separator}: ":"}
\end{Highlighting}
\end{Shaded}

According to the previous definition, each element has the size 3, thus it does not matter that the separator is possibly included in the elements. For example the string
\texttt{001:0:::002:2:1:112::::} would be parsed into the six elements \texttt{001},
\texttt{0::}, \texttt{002}, \texttt{2:1}, \texttt{112} and \texttt{:::}.

\subsubsection*{Enclosing strings}

In some cases, constant strings are present before the first and/of after the last element of a list, for example an opening and a closing bracket:

\begin{Shaded}
\begin{Highlighting}[]
\SectionTok{datatypes}\KeywordTok{:}
\AttributeTok{  }\DtnameTok{list7}\KeywordTok{:}
\AttributeTok{    }\DefkeyTok{list\_of}\KeywordTok{:}\AttributeTok{ unsigned\_integer}
\AttributeTok{    }\OptionTok{splitted\_by}\KeywordTok{:}\AttributeTok{ }\StringTok{","}
\AttributeTok{    }\OptionTok{prefix}\KeywordTok{:}\AttributeTok{ }\StringTok{"("}
\AttributeTok{    }\OptionTok{suffix}\KeywordTok{:}\AttributeTok{ }\StringTok{")"}
\end{Highlighting}
\end{Shaded}

This definition allows to parse a string representation like \texttt{(1,2,3,4)}.

\subsubsection*{Number of elements}

In some cases there is a minimum and/or maximum or a fixed number
elements of the list. This can be enforced using validation rules, as in the following examples:

\begin{Shaded}
\begin{Highlighting}[]
\SectionTok{datatypes}\KeywordTok{:}
\AttributeTok{  }\DtnameTok{list8}\KeywordTok{:}\CommentTok{   \# e.g. 0;{-}1;32}
\AttributeTok{    }\DefkeyTok{list\_of}\KeywordTok{:}\AttributeTok{ integer}
\AttributeTok{    }\OptionTok{splitted\_by}\KeywordTok{:}\AttributeTok{ }\StringTok{";"}
\AttributeTok{    }\FunctionTok{length}\KeywordTok{:}\AttributeTok{ }\DecValTok{3}
\AttributeTok{  }\FunctionTok{list9}\KeywordTok{:}
\AttributeTok{    }\DefkeyTok{list\_of}\KeywordTok{:}\AttributeTok{ integer}
\AttributeTok{    }\OptionTok{splitted\_by}\KeywordTok{:}\AttributeTok{ }\StringTok{";"}
\AttributeTok{    }\OptionTok{min\_length}\KeywordTok{:}\AttributeTok{ }\DecValTok{5}
\AttributeTok{    }\OptionTok{max\_length}\KeywordTok{:}\AttributeTok{ }\DecValTok{7}
\end{Highlighting}
\end{Shaded}

\subsubsection*{Empty lists}

Empty lists are also supported. If there is a prefix and suffix, the empty list is recognizable and does not require a special handling. If there are no enclosing strings, then the representation of an empty list is an empty string. This case can be handled, by using the \texttt{empty} option:

\begin{Shaded}
\begin{Highlighting}[]
\SectionTok{datatypes}\KeywordTok{:}
\AttributeTok{  }\DtnameTok{list9}\KeywordTok{:}
\AttributeTok{    }\DefkeyTok{list\_of}\KeywordTok{:}\AttributeTok{ }\KeywordTok{\{}\DefkeyTok{regex}\KeywordTok{:}\AttributeTok{ }\StringTok{"[A{-}Z]"}\KeywordTok{\}}
\AttributeTok{    }\OptionTok{empty}\KeywordTok{:}\AttributeTok{ }\KeywordTok{[]}
\AttributeTok{\textasciigrave{}\textasciigrave{}}
\end{Highlighting}
\end{Shaded}

\subsubsection*{Predefined representations}

Using decoding mappings and/or a default decoded value (see below) is it
possible to decode given strings to predefined lists.
In case multiple representations of the same value are given, the
\texttt{canonical} option must define which one shall be used for encoding, as in the following example:

\begin{Shaded}
\begin{Highlighting}[]
\SectionTok{datatypes}\KeywordTok{:}
\AttributeTok{  }\DtnameTok{list10}\KeywordTok{:}
\AttributeTok{    }\DefkeyTok{values}\KeywordTok{:}
\AttributeTok{      }\FunctionTok{"a"}\KeywordTok{:}\AttributeTok{ }\KeywordTok{[}\StringTok{"a"}\KeywordTok{]}
\AttributeTok{      }\FunctionTok{"1a"}\KeywordTok{:}\AttributeTok{ }\KeywordTok{[}\StringTok{"a"}\KeywordTok{]}
\AttributeTok{      }\FunctionTok{"2a"}\KeywordTok{:}\AttributeTok{ }\KeywordTok{[}\StringTok{"a"}\KeywordTok{,}\AttributeTok{ }\StringTok{"a"}\KeywordTok{]}
\AttributeTok{      }\FunctionTok{"3a"}\KeywordTok{:}\AttributeTok{ }\KeywordTok{[}\StringTok{"a"}\KeywordTok{,}\AttributeTok{ }\StringTok{"a"}\KeywordTok{,}\AttributeTok{ }\StringTok{"a"}\KeywordTok{]}
\AttributeTok{    }\OptionTok{empty}\KeywordTok{:}\AttributeTok{ }\KeywordTok{[]}
\AttributeTok{    }\OptionTok{canonical}\KeywordTok{:}\AttributeTok{ }\KeywordTok{\{}\FunctionTok{"1a"}\KeywordTok{:}\AttributeTok{ }\KeywordTok{[}\StringTok{"a"}\KeywordTok{]\}}
\end{Highlighting}
\end{Shaded}

\vfill\null
\columnbreak

\subsubsection*{Sets and Multisets}

Sometimes a different kind of collection is available in case the order of the elements is not important, and different kind of data structures are used for representing them in memory, such as hash tables. This kind of collections are available as e.g. \textit{sets} in Python and \textit{hash sets} in Nim.

There is no special handling for sets in TextFormats.
This means that the elements of sets are regarded as a list.
Thus, equivalence operations which disregard the order of the elements
must be implemented externally. Similarly, the uniqueness of the set elements
(vs.\ multisets) must be validated externally.

\subsubsection*{Heterogeneous lists}

A \texttt{one\_of} definition can be combined with a \texttt{list\_of} definition to implement lists of elements which have different types, which is not dependent on the positional order of the element and is not explicitly annotated by a key or typecode.
Instead, it is the formatting of the element itself which reveals the type.
For example, the following defines a list containing either
integers or single upcase characters:

\begin{Shaded}
\begin{Highlighting}[]
\SectionTok{datatypes}\KeywordTok{:}
\AttributeTok{  }\DtnameTok{list11}\KeywordTok{:}
\AttributeTok{    }\DefkeyTok{list\_of}\KeywordTok{:}
\AttributeTok{      }\DefkeyTok{one\_of}\KeywordTok{:}
\AttributeTok{        }\KeywordTok{{-}}\DefkeyTok{ integer}
\AttributeTok{        }\KeywordTok{{-}}\DefkeyTok{ regex}\KeywordTok{:}\AttributeTok{ }\FunctionTok{"[A-Z]"}
\
\AttributeTok{   }\OptionTok{splitted\_by}\KeywordTok{:}\AttributeTok{ }\StringTok{","}
\end{Highlighting}
\end{Shaded}

An example of string which can be handled by the previous definition is:
\texttt{1,{-}3,A,5,B,{-}2}.

\subsection*{Mappings, Dictionaries, Associative arrays}

In another type of compound values, the semantics and data type of the elements can differ, and different instances may contain or not some of the elements, or sometimes contain multiple elements of the same type. Such compound values are usually stored in open associative data structures, with different names and underlying data structures, such as \textit{dictionaries} (Python), \textit{tables} (Nim), \textit{hash tables} (C), \textit{objects} (JSON) or \textit{maps} (YAML).

In a text format, this kind of data can be represented in different ways. For example, the semantic and data type can be specified explicitly in the text representation, or be implicit by the format of the text itself. Depending on this, the kind of definition to be used in TextFormats differs (e.g. \texttt{list\_of} with \texttt{one\_of} elements, \texttt{labeled\_list} or \texttt{tagged\_list}).

\subsubsection{Semantics by format}

If the semantics of the elements of a list is determined by the format, a
\texttt{list\_of} definition can be given, in which the element is defined
using a \texttt{one\_of} definition. This is the same case illustrated above
under the paragraph \textit{heterogeneous lists}. In this case, which does not occur very often in practice, the result of parsing and the data to be passed to the encoding function must be a list. Thus some external preprocessing or postprocessing will be necessary to transform the data to or from a mapping.

\subsubsection{Key/value pairs}

In many cases, a collection contains elements of different type, and the semantics of each elements is given explicitly, alongside the value of the element. Thus, each element is present as a tuple of keys and values. Although a \texttt{list\_of} could
be used also for this case, TextFormats offers a specialized kind of list definition
for it, in case the set of possible keys and their associated data types are
known in advance.

In such cases a \texttt{labeled\_list} definition is used, as in the following example:

\begin{Shaded}
\begin{Highlighting}[]
\SectionTok{datatypes}\KeywordTok{:}
\AttributeTok{  }\DtnameTok{map1}\KeywordTok{:}
\AttributeTok{    }\DefkeyTok{labeled\_list}\KeywordTok{:}
\AttributeTok{      }\FunctionTok{rank}\KeywordTok{:}\AttributeTok{ unsigned\_integer}
\AttributeTok{      }\FunctionTok{name}\KeywordTok{:}\AttributeTok{ string}
\AttributeTok{    }\OptionTok{splitted\_by}\KeywordTok{:}\AttributeTok{ }\StringTok{";"}
\end{Highlighting}
\end{Shaded}

The set of possible keys and the datatypes of the values for each of the keys
are given under the \texttt{labeled\_list} key, as a mapping.
An \texttt{internal\_separator} string can be specified,
separating the key from the value (the default is \texttt{:}). The
internal separator cannot be empty and cannot occur in the keys, but it
can occur in the values. This condition is generally met in formats which
implement key/values lists.

\subsubsection{Single-instance keys}
Names are by default allowed to present multiple times in the list. For
this reason, the elements values are always given in the decoded value
as lists. In some cases, all or some of names can only be present once.
This can be enforced by listing them under the \texttt{single} key:

\begin{Shaded}
\begin{Highlighting}[]
\SectionTok{datatypes}\KeywordTok{:}
\AttributeTok{  }\DtnameTok{map2}\KeywordTok{:}
\AttributeTok{    }\DefkeyTok{labeled\_list}\KeywordTok{:}
\AttributeTok{      }\FunctionTok{rank}\KeywordTok{:}\AttributeTok{ unsigned\_integer}
\AttributeTok{      }\FunctionTok{name}\KeywordTok{:}\AttributeTok{ string}
\AttributeTok{    }\OptionTok{splitted\_by}\KeywordTok{:}\AttributeTok{ }\StringTok{";"}
\AttributeTok{    }\OptionTok{single}\KeywordTok{:}\AttributeTok{ }\KeywordTok{[}\AttributeTok{rank}\KeywordTok{,}\AttributeTok{ name}\KeywordTok{]}
\end{Highlighting}
\end{Shaded}

\subsubsection{Required keys}
Also, by default, some names may be absent in the set of elements. If
some of the names must be present, they are listed under the
\texttt{required} key:

\begin{Shaded}
\begin{Highlighting}[]
\SectionTok{datatypes}\KeywordTok{:}
\AttributeTok{  }\DtnameTok{map3}\KeywordTok{:}
\AttributeTok{    }\DefkeyTok{labeled\_list}\KeywordTok{:}
\AttributeTok{      }\FunctionTok{rank}\KeywordTok{:}\AttributeTok{ unsigned\_integer}
\AttributeTok{      }\FunctionTok{name}\KeywordTok{:}\AttributeTok{ string}
\AttributeTok{    }\OptionTok{splitted\_by}\KeywordTok{:}\AttributeTok{ }\StringTok{";"}
\AttributeTok{    }\OptionTok{internal\_separator}\KeywordTok{:}\AttributeTok{ }\StringTok{"="}
\AttributeTok{    }\OptionTok{required}\KeywordTok{:}\AttributeTok{ }\KeywordTok{[}\AttributeTok{name}\KeywordTok{]}
\end{Highlighting}
\end{Shaded}

\subsubsection{SAM-style tags}

In some cases the values of a collection are
 each accompanied by a name and typecode, i.e.~as triples
value/name/typecode. A prominent example for this are SAM-style tags, which
often included in recent bioinformatics formats, e.g. GFA \citep{gfa1,gfa2} and VCF \citep{vcf}, after their
original definition for use in the SAM format \citep{sam}.


The difference with the key/value case is that
the name defines the semantic of the value, but not all names
(differently from labeled values lists) must be defined in advance.
Since the name is not necessarily predefined, the type must be explicitly
given, thus a typecode is present in the text representation. Each typecode
is associated to a datatype definition.

For these case, the \texttt{tagged\_list} definition key is used, under
which all available datatypes codes and the associated datatype
definitions are given under the \texttt{tagged\_list} key as a mapping.

An example of tagged list definition is given here:

\begin{Shaded}
\begin{Highlighting}[]
\SectionTok{datatypes}\KeywordTok{:}
\AttributeTok{  }\DtnameTok{map4}\KeywordTok{:}
\AttributeTok{    }\DefkeyTok{tagged\_list}\KeywordTok{:}
\AttributeTok{      }\FunctionTok{i}\KeywordTok{:}\AttributeTok{ integer}
\AttributeTok{      }\FunctionTok{f}\KeywordTok{:}\AttributeTok{ float}
\AttributeTok{    }\OptionTok{tagname}\KeywordTok{:}\AttributeTok{ }\StringTok{"[A{-}Z]"}
\AttributeTok{    }\OptionTok{internal\_separator}\KeywordTok{:}\AttributeTok{ }\StringTok{"."}
\AttributeTok{    }\OptionTok{splitted\_by}\KeywordTok{:}\AttributeTok{ }\StringTok{";"}
\end{Highlighting}
\end{Shaded}

The definition given above can e.g. handle the string representation
\texttt{A.i.12;B.f.1.3}, which is parsed into the two elements
\texttt{A} with the value 12 and \texttt{B} with the value 1.3.

\subsubsection{Generalized tags}
The valid names and their formatting is specified using a regular
expression. The internal separator key has a default value (colon, \texttt{:})
and it must be a non-empty string. It cannot be present in tagnames and
type codes (but can be present in values). This restriction is reasonable and e.g.\ met in the in SAM tags.

\subsubsection{Predefined representations}

Using decoding mappings and/or a default decoded value (see below) is it
possible to decode given strings to predefined mappings/dictionaries.
In case a data value has multiple string representations, the canonical
one must be specified, which is then used for encoding.

\begin{Shaded}
\begin{Highlighting}[]
\SectionTok{datatypes}\KeywordTok{:}
\AttributeTok{  }\DtnameTok{map5}\KeywordTok{:}
\AttributeTok{    }\DefkeyTok{values}\KeywordTok{:}
\AttributeTok{      }\FunctionTok{"ax"}\KeywordTok{:}\AttributeTok{ }\KeywordTok{\{}\FunctionTok{"a"}\KeywordTok{:}\AttributeTok{ }\StringTok{"x"}\KeywordTok{,}\AttributeTok{ }\FunctionTok{"b"}\KeywordTok{:}\AttributeTok{ }\StringTok{"y"}\KeywordTok{\}}
\AttributeTok{      }\FunctionTok{"a"}\KeywordTok{:}\AttributeTok{ }\KeywordTok{\{}\FunctionTok{"a"}\KeywordTok{:}\AttributeTok{ }\StringTok{"x"}\KeywordTok{,}\AttributeTok{ }\FunctionTok{"b"}\KeywordTok{:}\AttributeTok{ }\StringTok{"y"}\KeywordTok{\}}
\AttributeTok{      }\FunctionTok{"ay"}\KeywordTok{:}\AttributeTok{ }\KeywordTok{\{}\StringTok{"a"}\KeywordTok{,}\AttributeTok{ }\StringTok{"y"}\KeywordTok{,}\AttributeTok{ }\FunctionTok{"b"}\KeywordTok{:}\AttributeTok{ }\StringTok{"y"}\KeywordTok{\}}
\AttributeTok{      }\FunctionTok{"bx"}\KeywordTok{:}\AttributeTok{ }\KeywordTok{\{}\FunctionTok{"a"}\KeywordTok{:}\AttributeTok{ }\StringTok{"x"}\KeywordTok{,}\AttributeTok{ }\FunctionTok{"b"}\KeywordTok{:}\AttributeTok{ }\StringTok{"x"}\KeywordTok{\}}
\AttributeTok{    }\OptionTok{canonical}\KeywordTok{:}
\AttributeTok{      }\FunctionTok{"ax"}\KeywordTok{:}\AttributeTok{ }\KeywordTok{\{}\FunctionTok{"a"}\KeywordTok{:}\AttributeTok{ }\StringTok{"x"}\KeywordTok{,}\AttributeTok{ }\FunctionTok{"b"}\KeywordTok{:}\AttributeTok{ }\StringTok{"y"}\KeywordTok{\}}
\AttributeTok{      }\FunctionTok{"ay"}\KeywordTok{:}\AttributeTok{ }\KeywordTok{\{}\StringTok{"a"}\KeywordTok{,}\AttributeTok{ }\StringTok{"y"}\KeywordTok{,}\AttributeTok{ }\FunctionTok{"b"}\KeywordTok{:}\AttributeTok{ }\StringTok{"y"}\KeywordTok{\}}
\AttributeTok{      }\FunctionTok{"bx"}\KeywordTok{:}\AttributeTok{ }\KeywordTok{\{}\FunctionTok{"a"}\KeywordTok{:}\AttributeTok{ }\StringTok{"x"}\KeywordTok{,}\AttributeTok{ }\FunctionTok{"b"}\KeywordTok{:}\AttributeTok{ }\StringTok{"x"}\KeywordTok{\}}
\AttributeTok{    }\OptionTok{empty}\KeywordTok{:}\AttributeTok{  }\KeywordTok{\{}\FunctionTok{"a"}\KeywordTok{:}\AttributeTok{ }\StringTok{"z"}\KeywordTok{,}\AttributeTok{ }\FunctionTok{"b"}\KeywordTok{:}\AttributeTok{ }\StringTok{"z"}\KeywordTok{\}}
\end{Highlighting}
\end{Shaded}

\subsubsection{Implicit entries}

In some cases, the decoded dictionary shall contain a given constant
key/value pair, which is not explicitely encoded. These are specified
using the \texttt{implicit} mapping, which is available for
\texttt{composed\_of}, \texttt{labeled\_list} and \texttt{tagged\_list}
definitions:

\begin{Shaded}
\begin{Highlighting}[]
\SectionTok{datatypes}\KeywordTok{:}
\AttributeTok{  }\DtnameTok{map6}\KeywordTok{:}
\AttributeTok{    }\DefkeyTok{composed\_of}\KeywordTok{:}
\AttributeTok{      }\KeywordTok{{-}}\AttributeTok{ }\FunctionTok{name}\KeywordTok{:}\AttributeTok{ string}
\AttributeTok{      }\KeywordTok{{-}}\AttributeTok{ }\FunctionTok{copies}\KeywordTok{:}\AttributeTok{ unsigned\_integer}
\AttributeTok{    }\OptionTok{splitted\_by}\KeywordTok{:}\AttributeTok{ }\StringTok{","}
\AttributeTok{    }\OptionTok{implicit}\KeywordTok{:}\AttributeTok{ }\KeywordTok{\{}\FunctionTok{type}\KeywordTok{:}\AttributeTok{ }\StringTok{"rRNA"}\KeywordTok{\}}
\end{Highlighting}
\end{Shaded}

This definition can e.g. handle the string representation \texttt{"16S,2"}, which is parsed into the mapping with 3 keys \texttt{\{"name": "16S", copies: 2, type: "rRNA"\}}.

\subsection*{Objects, Structs}

In this section we handle collections of values where each instance contain the same set of elements (with possible exceptions), representing different aspects of the data. Each of the element has its own data type and semantics.

This kind of compound values is represented in \textit{structs} (C, Python) or instances of \textit{classes} (Python, Nim). Other possible representations are those mentioned in the previous paragraph (mappings, hash tables), eventually adding validations, making sure that all and only the correct elements are present.

In TextFormats the description of this kind of data is perfomed using \texttt{composed\_of} definitions. Under the definition key,
a list is of tuples is given, each one as name and
definition. Note that this is a list (thus the \texttt{-} in YAML) and
not a mapping, since the order of the elements is important, as it defines which element is which:

\begin{Shaded}
\begin{Highlighting}[]
\SectionTok{datatypes}\KeywordTok{:}
\AttributeTok{  }\DtnameTok{obj1}\KeywordTok{:}
\AttributeTok{    }\DefkeyTok{composed\_of}\KeywordTok{:}
\AttributeTok{      }\KeywordTok{{-}}\AttributeTok{ }\FunctionTok{first}\KeywordTok{:}\AttributeTok{ unsigned\_integer}
\AttributeTok{      }\KeywordTok{{-}}\AttributeTok{ }\FunctionTok{second}\KeywordTok{:}\AttributeTok{ float}
\AttributeTok{      }\KeywordTok{{-}}\AttributeTok{ }\FunctionTok{third}\KeywordTok{:}\AttributeTok{ }\KeywordTok{\{}\DefkeyTok{regex}\KeywordTok{:}\AttributeTok{ }\StringTok{"[A{-}Za{-}z0{-}9]"}\KeywordTok{\}}
\AttributeTok{    }\OptionTok{splitted\_by}\KeywordTok{:}\AttributeTok{ }\StringTok{" "}
\end{Highlighting}
\end{Shaded}

\subsubsection*{Enclosing strings}

As for lists, \texttt{composed\_of} definitions can include a \texttt{prefix} and/or a \texttt{suffix} option, which define enclosing strings (e.g. parentheses) before the first element and/or after the last element.

\subsubsection*{Element separators}

The \texttt{splitted\_by} and \texttt{separator} options are used for describing how to separate the single elements of the compound value. These have the same kind of usage already illustrated in the Lists section, i.e. \texttt{splitted\_by} is used for separators which cannot occur in the elements text, while \texttt{separator} is used
otherwise, e.g. when the escaped separator can occur in the elements or the element size is recognized by their format, e.g. fixed-length elements.

\subsubsection*{Multiple separators}

In some cases, different separators are used between different pairs of
elements. In this case, they can be specified as additional constant elements and
hidden in the decoded dictionary using the \texttt{hide\_constants} option, as in the following example:

\begin{Shaded}
\begin{Highlighting}[]
\SectionTok{datatypes}\KeywordTok{:}
\AttributeTok{  }\DtnameTok{obj2}\KeywordTok{:}
\AttributeTok{    }\DefkeyTok{composed\_of}\KeywordTok{:}
\AttributeTok{      }\KeywordTok{{-}}\AttributeTok{ }\FunctionTok{x}\KeywordTok{:}\DefkeyTok{ unsigned\_integer}
\AttributeTok{      }\KeywordTok{{-}}\AttributeTok{ }\FunctionTok{sep1}\KeywordTok{:}\AttributeTok{ }\KeywordTok{\{}\DefkeyTok{constant}\KeywordTok{:}\AttributeTok{ }\StringTok{";"}\KeywordTok{\}}
\AttributeTok{      }\KeywordTok{{-}}\AttributeTok{ }\FunctionTok{y}\KeywordTok{:}\DefkeyTok{ float}
\AttributeTok{      }\KeywordTok{{-}}\AttributeTok{ }\FunctionTok{sep2}\KeywordTok{:}\AttributeTok{ }\KeywordTok{\{}\DefkeyTok{constant}\KeywordTok{:}\AttributeTok{ }\StringTok{"|"}\KeywordTok{\}}
\AttributeTok{      }\KeywordTok{{-}}\AttributeTok{ }\FunctionTok{z}\KeywordTok{:}\AttributeTok{ }\KeywordTok{\{}\DefkeyTok{regex}\KeywordTok{:}\AttributeTok{ }\StringTok{"[A{-}Za{-}z]"}\KeywordTok{\}}
\AttributeTok{    }\OptionTok{hide\_constants}\KeywordTok{:}\AttributeTok{ }\CharTok{true}
\end{Highlighting}
\end{Shaded}

An example of string representation which is parsed by the previous definition is \texttt{1;2.0|A}. Hereby the resulting mapping has three elements, as the constant separators are only considered for parsing: \texttt{\{x:1, y:2.0, z:A\}}.

\subsubsection*{Optional separated elements}

Some of the elements can be optional, i.e.~sometimes absent from the
sequence of elements. In case the sequence is splitted by a non-empty separator
  string, an empty element can be recognized by the presence of
  this separator. In this case the \texttt{empty} option is used in the
  elements definitions, as in the following example:

\begin{Shaded}
\begin{Highlighting}[]
\SectionTok{datatypes}\KeywordTok{:}
\AttributeTok{  }\DtnameTok{obj3}\KeywordTok{:}
\AttributeTok{    }\DefkeyTok{composed\_of}\KeywordTok{:}
\AttributeTok{      }\KeywordTok{{-}}\AttributeTok{ }\FunctionTok{first}\KeywordTok{:}\AttributeTok{ }\KeywordTok{\{}\DefkeyTok{values}\KeywordTok{:}\AttributeTok{ }\KeywordTok{[}\DecValTok{1}\KeywordTok{,}\DecValTok{2}\KeywordTok{],}\AttributeTok{ }\OptionTok{empty}\KeywordTok{:}\AttributeTok{ }\DecValTok{0}\KeywordTok{\}}
\AttributeTok{      }\KeywordTok{{-}}\AttributeTok{ }\FunctionTok{second}\KeywordTok{:}\AttributeTok{ }\KeywordTok{\{}\DefkeyTok{values}\KeywordTok{:}\AttributeTok{ }\KeywordTok{[}\AttributeTok{A}\KeywordTok{,}\AttributeTok{B}\KeywordTok{],}\AttributeTok{ }\FunctionTok{empty}\KeywordTok{:}\AttributeTok{ C}\KeywordTok{\}}
\AttributeTok{    }\OptionTok{splitted\_by}\KeywordTok{:}\AttributeTok{ }\StringTok{";"}
\end{Highlighting}
\end{Shaded}

\subsubsection*{Optional trailing elements}

  In some cases, a given number of elements at the beginning of the text representation
  may be mandatory, while the following can be present or not (and are
  mandatory only if elements after them in the order are present). In
  this case, the \texttt{required} option is used (number of required elements):

\begin{Shaded}
\begin{Highlighting}[]
\SectionTok{datatypes}\KeywordTok{:}
\AttributeTok{  }\DtnameTok{obj4}\KeywordTok{:}
\AttributeTok{    }\DefkeyTok{composed\_of}\KeywordTok{:}
\AttributeTok{      }\KeywordTok{{-}}\AttributeTok{ }\FunctionTok{first}\KeywordTok{:}\AttributeTok{ }\KeywordTok{\{}\DefkeyTok{values}\KeywordTok{:}\AttributeTok{ }\KeywordTok{[}\DecValTok{1}\KeywordTok{,}\DecValTok{2}\KeywordTok{],}\AttributeTok{ }\OptionTok{empty}\KeywordTok{:}\AttributeTok{ }\DecValTok{0}\KeywordTok{\}}
\AttributeTok{      }\KeywordTok{{-}}\AttributeTok{ }\FunctionTok{second}\KeywordTok{:}\AttributeTok{ }\KeywordTok{\{}\DefkeyTok{values}\KeywordTok{:}\AttributeTok{ }\KeywordTok{[}\AttributeTok{A}\KeywordTok{,}\AttributeTok{B}\KeywordTok{],}\AttributeTok{ }\OptionTok{empty}\KeywordTok{:}\AttributeTok{ C}\KeywordTok{\}}
\AttributeTok{    }\OptionTok{splitted\_by}\KeywordTok{:}\AttributeTok{ }\StringTok{";"}
\AttributeTok{    }\OptionTok{required}\KeywordTok{:}\AttributeTok{ }\DecValTok{1}
\end{Highlighting}
\end{Shaded}

\subsubsection*{Optional internal elements}

  In the case, some internal element can be missing (together with
  its associated separator, or when no separator is used) but there is
  no ambuiguity,
  e.g.~because the following element of the sequence has a type that
  allows it to be distinguished from the optional element, or because
  the total number of elements changes depending on the presence or
  absence of the optional element.

In this case the user must provide multiple alternative
definitions of the structure (i.e.~with and without the optional
element) using a \texttt{one\_of} definition.
Furthermore, in order to provide the same set of values for all instances of the object
or struct, the \texttt{implicit} option can be used.


\begin{Shaded}
\begin{Highlighting}[]
\SectionTok{datatypes}\KeywordTok{:}
\AttributeTok{  }\DtnameTok{obj5}\KeywordTok{:}
\AttributeTok{    }\DefkeyTok{one\_of}\KeywordTok{:}
\AttributeTok{    }\KeywordTok{{-}}\AttributeTok{ }\DefkeyTok{composed\_of}\KeywordTok{:}
\AttributeTok{        }\KeywordTok{{-}}\AttributeTok{ }\FunctionTok{name}\KeywordTok{:}\AttributeTok{ string}
\AttributeTok{      }\KeywordTok{{-}}\AttributeTok{ }\FunctionTok{expressed}\KeywordTok{:}
\AttributeTok{          }\DefkeyTok{values}\KeywordTok{:}\AttributeTok{ }\KeywordTok{\{}\FunctionTok{"+"}\KeywordTok{:}\AttributeTok{ }\CharTok{true}\KeywordTok{,}\AttributeTok{ }\FunctionTok{"{-}"}\KeywordTok{:}\AttributeTok{ }\CharTok{false}\KeywordTok{\}}
\AttributeTok{     }\OptionTok{splitted\_by}\KeywordTok{:}\AttributeTok{ }\StringTok{","}
\AttributeTok{     }\OptionTok{implicit}: \KeywordTok{\{}\StringTok{"copies"}\KeywordTok{:}\AttributeTok{ 1}\KeywordTok{\}}
\AttributeTok{    }\KeywordTok{{-}}\AttributeTok{ }\DefkeyTok{composed\_of}\KeywordTok{:}
\AttributeTok{      }\KeywordTok{{-}}\AttributeTok{ }\FunctionTok{name}\KeywordTok{:}\DefkeyTok{ string}
\AttributeTok{      }\KeywordTok{{-}}\AttributeTok{ }\FunctionTok{copies}\KeywordTok{:}\DefkeyTok{ unsigned\_integer}
\AttributeTok{      }\KeywordTok{{-}}\AttributeTok{ }\FunctionTok{expressed}\KeywordTok{:}
\AttributeTok{          }\DefkeyTok{values}\KeywordTok{:}\AttributeTok{ }\KeywordTok{\{}\FunctionTok{"+"}\KeywordTok{:}\AttributeTok{ }\CharTok{true}\KeywordTok{,}\AttributeTok{ }\FunctionTok{"{-}"}\KeywordTok{:}\AttributeTok{ }\CharTok{false}\KeywordTok{\}}
\AttributeTok{      }\OptionTok{splitted\_by}\KeywordTok{:}\AttributeTok{ }\StringTok{","}
\end{Highlighting}
\end{Shaded}

The above definition would parse the string representation
\texttt{"X,+"} to the mapping
\texttt{\{name: X, expressed: true, copies: 1\}}, while \texttt{"X,2,+"} would
be parsed to \texttt{\{name: X, expressed: true, copies: 2\}}.

\subsection*{Unions}

In some cases an element of a format can be expressed in multiple
different ways. In dynamically typed languages such as Python, any variable can store this kind of values. In C, such values could be e.g. stored as \textit{unions},
and in Nim as \textit{variant objects}.

In TextFormats the type of such values can be defined defined using definitions
of type  \texttt{one\_of}. E.g. the following allows to represent an unsigned integer value, as a number, if it is \textgreater= 1, otherwise a floating point:

\begin{Shaded}
\begin{Highlighting}[]
\SectionTok{datatypes}\KeywordTok{:}
\AttributeTok{  }\DtnameTok{num10}\KeywordTok{:}
\AttributeTok{    }\FunctionTok{one\_of}\KeywordTok{:}
\AttributeTok{      }\KeywordTok{{-}}\AttributeTok{ }\DefkeyTok{unsigned\_integer}\KeywordTok{:}\AttributeTok{ }\KeywordTok{\{}\OptionTok{min}\KeywordTok{:}\AttributeTok{ }\DecValTok{1}\KeywordTok{\}}
\AttributeTok{      }\KeywordTok{{-}}\AttributeTok{ }\DefkeyTok{float}
\end{Highlighting}
\end{Shaded}

Note that the content of the \texttt{one\_of} key is a YAML list; the order of
the elements in the
list defines the order or precedence of the definitions (the first which
applies is used).

\section*{Conclusions}

In this paper, the representation of different kind of data in text formats, as specified using the library TextFormats, has been analysed. Thereby it was
demonstrated that most types of values that can be used in programming languages
such as C and Python, can also be represented in TextFormats.

In TextFormats specifications, the same type of value is sometimes represented using a different kind of datatype definition.
This is true for both scalar values and compound values. For example,
in the above examples, boolean values are sometimes represented using definitions
of kind 
\texttt{values}, sometimes \texttt{regexes} or even \texttt{constant}, depending on what is their representation in the format. Among the examples for compound values,
e.g., collections of tagged elements are sometimes represented using \texttt{list\_of}, but in other cases using the specialized list definition keys \texttt{tagged\_list} and \texttt{labeled\_list}.

The reason for this is that the TextFormat syntax for datatype definitions is oriented to the text representation and not to the type of represented value.
The systematic review of the definition types based on the type of value is particularly useful when defining a new format and complements the TFSL syntax manual included in the library
documentation, which is more useful when a specification is written for a format which already exists.

\bibliography{references}

\begin{thebibliography}{5}
\providecommand{\natexlab}[1]{#1}
\providecommand{\url}[1]{\texttt{#1}}
\expandafter\ifx\csname urlstyle\endcsname\relax
  \providecommand{\doi}[1]{doi: #1}\else
  \providecommand{\doi}{doi: \begingroup \urlstyle{rm}\Url}\fi

\bibitem[Gonnella(2022)]{textformats}
Giorgio Gonnella.
\newblock Textformats: Simplifying the definition and parsing of text formats
  in bioinformatics.
\newblock \emph{PLOS ONE}, 17\penalty0 (5):\penalty0 1--17, 05 2022.
\newblock \doi{10.1371/journal.pone.0268910}.
\newblock URL \url{https://doi.org/10.1371/journal.pone.0268910}.

\bibitem[{GFA Format Specification Working Group}(2016)]{gfa1}
{GFA Format Specification Working Group}.
\newblock The {G}{F}{A} format specification, 2016.
\newblock URL \url{http://gfa-spec.github.io/GFA-spec/GFA1.html}.

\bibitem[{GFA Format Specification Working Group}(2018)]{gfa2}
{GFA Format Specification Working Group}.
\newblock Graphical fragment assembly ({G}{F}{A}) 2.0 format specification,
  2018.
\newblock URL \url{http://gfa-spec.github.io/GFA-spec/GFA2.html}.

\bibitem[Danecek et~al.(2011)Danecek, Auton, Abecasis, Albers, Banks, DePristo,
  Handsaker, Lunter, Marth, Sherry, McVean, Durbin, and Group]{vcf}
Petr Danecek, Adam Auton, Goncalo Abecasis, Cornelis~A. Albers, Eric Banks,
  Mark~A. DePristo, Robert~E. Handsaker, Gerton Lunter, Gabor~T. Marth,
  Stephen~T. Sherry, Gilean McVean, Richard Durbin, and 1000 Genomes
  Project~Analysis Group.
\newblock {The variant call format and VCFtools}.
\newblock \emph{Bioinformatics}, 27\penalty0 (15):\penalty0 2156--2158, 06
  2011.
\newblock ISSN 1367-4803.
\newblock \doi{10.1093/bioinformatics/btr330}.
\newblock URL \url{https://doi.org/10.1093/bioinformatics/btr330}.

\bibitem[Li et~al.(2009)Li, Handsaker, Wysoker, Fennell, Ruan, Homer, Marth,
  Abecasis, Durbin, and Subgroup]{sam}
Heng Li, Bob Handsaker, Alec Wysoker, Tim Fennell, Jue Ruan, Nils Homer, Gabor
  Marth, Goncalo Abecasis, Richard Durbin, and 1000 Genome Project
  Data~Processing Subgroup.
\newblock {The Sequence Alignment/Map format and SAMtools}.
\newblock \emph{Bioinformatics}, 25\penalty0 (16):\penalty0 2078--2079, 06
  2009.
\newblock ISSN 1367-4803.
\newblock \doi{10.1093/bioinformatics/btp352}.
\newblock URL \url{https://doi.org/10.1093/bioinformatics/btp352}.

\end{thebibliography}

\begin{acknowledgements}
Giorgio Gonnella has been supported by the DFG Grant GO 3192/1-1 ‘`Automated characterization of microbial genomes and metagenomes by collection and verification of association rules’’. The funders had no role in study design, data collection and analysis, decision to publish, or preparation of the manuscript.
\end{acknowledgements}

\begin{contributions}
 These contributions follow the Contributor Roles Taxonomy guidelines: \href{https://casrai.org/credit/}{https://casrai.org/credit/}.
 Conceptualization: G.G.;
 Data curation: G.G.;
 Formal analysis:  G.G.;
 Funding acquisition:  G.G.;
 Investigation: G.G.;
 Methodology: G.G.;
 Project administration: G.G.;
 Resources: G.G.;
 Software: G.G.;
 Supervision: G.G.;
 Validation: G.G.;
 Visualization:  G.G.;
 Writing – original draft: G.G.;
 Writing – review \& editing: G.G.
\end{contributions}

\begin{interests}
 The authors declare no competing financial interests.
\end{interests}

\end{multicols}
\end{document}